\documentclass[10pt,a4paper,onecolumn]{article}

\usepackage[T1]{fontenc}
\usepackage{newtxtext,newtxmath}
\usepackage[margin=22mm,headheight=14pt,footskip=18pt]{geometry}
\usepackage{microtype}
\usepackage{setspace}
\usepackage{authblk}

\usepackage{amsmath,amsfonts,bm}
\usepackage{slashed,graphicx,float}
\usepackage{multirow,multicol,tabulary}
\usepackage[flushleft]{threeparttable}
\usepackage{pgfplots,subfigure,tikz,comment,environ,letltxmacro}
\usepackage[switch*,displaymath,pagewise,mathlines]{lineno}
\usepackage{cite,titlesec,fancyhdr,xcolor,nameref}
\definecolor{acsblue}{RGB}{17,76,139}
\definecolor{orng}{RGB}{0,140,60}
\definecolor{sari}{RGB}{214,170,0}
\definecolor{shadecolor}{RGB}{255,241,204}
\usepackage[pagebackref=false,colorlinks=true,linkcolor=acsblue,citecolor=red,urlcolor=acsblue,pdfstartview=FitH,bookmarksnumbered=true]{hyperref}

\titleformat{\section}{\large\bfseries\color{black}}{\thesection.}{0.55em}{}
\titleformat{\subsection}{\normalsize\bfseries\color{black}}{\thesubsection}{0.6em}{}
\titleformat{\subsubsection}{\normalsize\itshape\color{black}}{\thesubsubsection}{0.6em}{}
\titlespacing*{\section}{0pt}{2.0ex plus .5ex minus .2ex}{0.8ex}
\titlespacing*{\subsection}{0pt}{1.5ex plus .4ex minus .2ex}{0.5ex}
\numberwithin{equation}{section}
\pgfplotsset{compat=1.18}
\usepgflibrary{arrows}
\usetikzlibrary{shapes.callouts}
\tikzset{level/.style={thick},connect/.style={dotted,red},notice/.style={draw,rectangle callout,callout relative pointer={#1}},label/.style={text width=2cm}}

\makeatletter
\let\oldr@@t\r@@t
\def\r@@t#1#2{%
  \setbox0=\hbox{$\oldr@@t#1{#2\,}$}\dimen0=\ht0
  \advance\dimen0-0.2\ht0
  \setbox2=\hbox{\vrule height\ht0 depth -\dimen0}%
  {\box0\lower0.4pt\box2}}
\LetLtxMacro{\oldsqrt}{\sqrt}
\renewcommand*{\sqrt}[2][\ ]{\oldsqrt[#1]{#2}}
\makeatother

\begin{document}

\newcommand{{\ri}}{{\rm{i}}}
\newcommand{{\Psibar}}{{\bar{\Psi}}}
\renewcommand{\rmdefault}{ptm}

\title{Quasinormal Ringing of a Schwarzschild Black Hole Surrounded by a King Dark Matter Halo: Higher-Order WKB Analysis}

\author{ \textit {\mdseries{Semra Gurtas Dogan}}$^{\ 1}$\footnote{\textit{ E-mail: semragurtasdogan@hakkari.edu.tr (\textcolor{acsblue}{Corresponding Author}) } }~,~ \textit {\mdseries{Hassan Hassanabadi}}$^{\ 2,3}$\footnote{\textit{ E-mail:hha1349@gmail.com  } }~,~ \textit {\mdseries{Abdullah Guvendi}}$^{\ 4}$\footnote{\textit{ E-mail: abdullah.guvendi@erzurum.edu.tr } }~   \\
	\small \textit {$^{\ 1}$ Department of Medical Imaging Techniques, Hakkari University, 30000, Hakkari, Türkiye}\\
	\small \textit {$^{\ 2}$  Physics Department, California State University, Fresno, CA 93740, USA} \\
 \small \textit {$^{\ 3}$ Department of Physics, Faculty of Science, University of Hradec Králové, Rokitanského 62, 500 03 Hradec Králové, Czechia}\\
 \small \textit {$^{\ 4}$Department of Basic Sciences, Erzurum Technical University, 25050, Erzurum,		Türkiye}}
	
\date{}
\maketitle

\begin{abstract}
We study scalar and electromagnetic perturbations of a Schwarzschild black hole (BH) surrounded by a dark matter halo described by the King density profile. The lapse function can be expressed as $1-2[M+M_{\rm DM}(r)]/r$, where $M_{\rm DM}(r)$ is the halo mass enclosed within radius $r$; thus, the halo affects the geometry through its enclosed-mass function. We compute the quasinormal frequencies using the sixth-order WKB method and its Pad\'e-improved version. Rewriting the lapse function in terms of the inverse hyperbolic sine avoids the severe numerical cancellations that arise when evaluating its high-order derivatives. For the configurations examined, the fundamental frequencies agree with those extracted from time-domain evolution to within $0.2\%$.The halo lowers both the oscillation frequency and the damping rate. The change of the real part is almost entirely accounted for by the halo mass enclosed by the photon sphere, and the additional decrease of the damping rate follows the ratio of the Lyapunov exponent to the angular frequency of the circular null orbit. Since part of the halo lies inside the horizon in this metric, we also place the halo outside the BH by means of an Einstein cluster. The frequencies then decrease by a similar amount, but the real part follows the Schwarzschild value multiplied by the gravitational redshift factor of the halo. A jump in the density at the inner edge of the halo scatters the waves and changes the damping rate by up to about $9\%$. For a continuous density profile, deviations from the redshift relation remain below about $1.5\%$ for $r_{\rm in}=3M$ and below about $0.5\%$ for the cases with $r_{\rm in}\geq4M$. No unstable modes are found within the perturbation sectors and parameter configurations examined.\\


\textbf{Keywords:} Black hole quasinormal modes; Dark matter halo; King density profile; Higher-order WKB approximation; Einstein cluster
\end{abstract}

\section{Introduction}
\label{sec:intro}
A perturbed BH relaxes through damped oscillations known as quasinormal modes (QNMs), whose complex frequencies are determined by the background geometry and by the boundary conditions at the horizon and at infinity. QNMs characterize the ringdown observed in gravitational-wave signals since 2015~\cite{LIGO2016}; together with BH shadow measurements~\cite{EHT2019}, they provide complementary probes of the strong-field region. The theory of QNMs and the methods used to compute them are reviewed in Refs.~\cite{Kokkotas1999,Berti2009,Konoplya2011}.

Real astrophysical BHs are not isolated: supermassive BHs reside in galactic environments that may include dark matter halos. A key question is how such environments modify the QNM spectrum. A common approach is to choose a density profile for the halo and to construct a static, spherically symmetric metric for the BH together with the halo~\cite{Xu2018}. Galactic BHs have also been described by exact solutions with an anisotropic fluid~\cite{Cardoso2022}. Here we are interested in the King profile~\cite{King1962},
\begin{equation}
\rho(r)=\rho_0\left(1+\frac{r^2}{R^2}\right)^{-3/2},
\label{eq:king}
\end{equation}
which has a core of nearly constant density and radius $R$ and falls off as $r^{-3}$ for $r\gg R$. The corresponding BH metric was given in Refs.~\cite{Kar2025,Hassanabadi2026}, and BHs in King halos were studied in several recent papers~\cite{Zare2026,AlBadawi2026}. Periodic orbits and scalar QNMs of this BH were obtained in Ref.~\cite{Hassanabadi2026} with the Mashhoon approximation and the third order WKB formula, using the single parameter $8\pi\rho_0R^2$ at fixed $M/R$, and QNMs of the rotating counterpart were computed in Ref.~\cite{Wang2026}. For other halo profiles, such as the Hernquist one, QNMs have been studied with similar methods~\cite{Qi2026}.

The present work complements Ref.~\cite{Hassanabadi2026}. We use the sixth-order WKB formula~\cite{Konoplya2003} and its Pad\'e-improved form~\cite{Matyjasek2017,Konoplya2019}, vary $\rho_0$ and $R$ independently, include both scalar and electromagnetic perturbations, and validate the frequency estimates against time-domain evolution~\cite{Gundlach1994}. The lapse function in its usual form turns out to be badly suited for the high derivatives required by the WKB formula, and we explain how this difficulty can be avoided. The tables of frequencies have a simple interpretation: the real part is close to that of a Schwarzschild BH with the mass enclosed by the photon sphere, while the damping rate is reduced further by an amount set by the circular null orbit. The strategy follows the approach used in Ref.~\cite{Pereira2026} for a Schwarzschild Klinkhamer wormhole. Taken together, the higher-order semi-analytic calculation, time-domain cross-check, and alternative matter construction test both the numerical reliability and the physical interpretation of the spectral shifts.

In the first metric, part of the halo lies inside the event horizon, making a literal interpretation as a static matter distribution problematic. The field equations are satisfied exactly, but the condition $g_{tt}g_{rr}=-1$ requires a radial tension, $P_r=-\rho$, and such a source cannot be interpreted as a static halo of collisionless dark matter. We therefore use this geometry as a phenomenological reference. In the second part of the paper, we instead place the King halo outside the BH using the Einstein-cluster construction~\cite{Einstein1939}, previously applied to a Hernquist halo by Cardoso \emph{et al.}~\cite{Cardoso2022} and to Dehnen halos with an inner edge by Hassanabadi and Zare~\cite{Hassanabadi2026b}. We then compute the exterior-halo spectrum, compare it with the phenomenological model, and investigate how the halo inner edge scatters the waves. Comparing the two configurations clarifies how the predicted spectral shifts depend on the assumed halo geometry and matter model. The comparison distinguishes the role of the enclosed halo mass in the phenomenological metric from the gravitational redshift and inner-edge scattering in the Einstein-cluster construction, thereby identifying which spectral trends persist when the halo is modeled using collisionless matter.

The paper is organized as follows. Section~\hyperref[sec:geometry]{2} introduces the BH-halo geometry and its parameters, while Sec.~\hyperref[sec:pert]{3} formulates the perturbation equations and describes the WKB and Pad\'e-improved calculations. Section~\hyperref[sec:results]{4} presents the quasinormal spectrum and its physical interpretation, and Sec.~\hyperref[sec:td]{5} tests the frequency estimates against time-domain evolution. Section~\hyperref[sec:ec]{6} develops the Einstein-cluster model with the halo outside the horizon and examines the effects of its inner edge. The conclusions are given in Sec.~\hyperref[sec:conc]{7}. We set $G=c=1$ throughout and use $M=1$ for all numerical results.

\section{Black hole in a King halo}
\label{sec:geometry}
We first specify the static, spherically symmetric geometry and identify how the King profile modifies the enclosed mass, horizon, and asymptotic behavior. We consider the line element
\begin{equation}
ds^2=-f(r)\,dt^2+\frac{dr^2}{f(r)}+r^2\left(d\theta^2+\sin^2\theta\,d\varphi^2\right),
\label{eq:metric}
\end{equation}
with the lapse function~\cite{Kar2025,Hassanabadi2026}
\begin{equation}
f(r)=1-\frac{2M}{r}+\frac{8\pi\rho_0R^3}{\sqrt{r^2+R^2}}+\frac{8\pi\rho_0R^3}{r}\ln\!\left(\frac{\sqrt{r^2+R^2}-r}{R}\right).
\label{eq:f}
\end{equation}
Since $\ln[(\sqrt{r^2+R^2}-r)/R]=-{\rm arcsinh}(r/R)$, this takes the form
\begin{equation}
f(r)=1-\frac{2\left[M+M_{\rm DM}(r)\right]}{r},\qquad
M_{\rm DM}(r)=4\pi\rho_0R^3\left[{\rm arcsinh}\frac{r}{R}-\frac{r}{\sqrt{r^2+R^2}}\right],
\label{eq:Mdm}
\end{equation}
and differentiation shows that $M_{\rm DM}(r)=\int_0^r4\pi s^2\rho(s)\,ds$ is the mass of the halo~\eqref{eq:king} inside the radius $r$. The halo thus acts as an additional mass that grows with the radius, $M_{\rm DM}\simeq\frac{4\pi}{3}\rho_0r^3$ inside the core and $M_{\rm DM}\simeq4\pi\rho_0R^3[\ln(2r/R)-1]$ well outside it. Its overall size is set by $\rho_0R^3$, so changing the core radius has a much stronger effect than a comparable relative change of the central density. For $\rho_0\to0$ or $R\to0$ the Schwarzschild metric is recovered.

\begin{figure}[htbp]
\centering
\includegraphics[width=\textwidth]{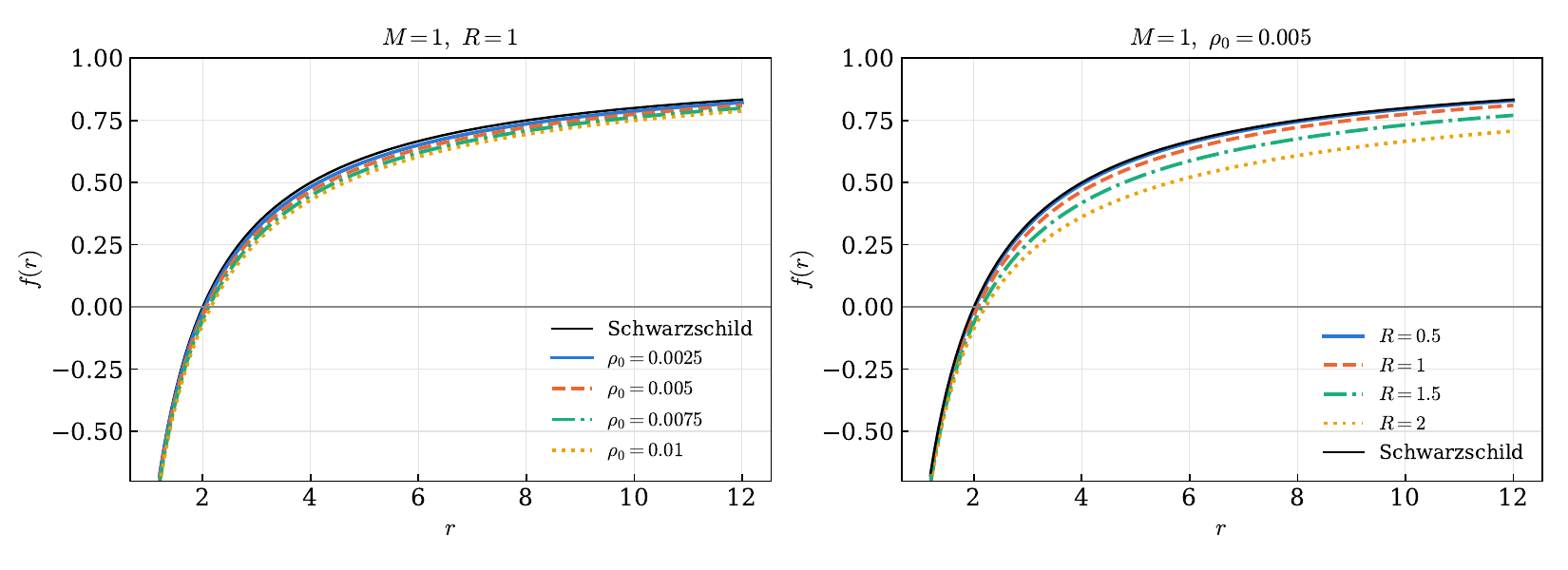}
\caption{The lapse function~\eqref{eq:f} for $M=1$. Left: $R=1$ and several values of $\rho_0$. Right: $\rho_0=0.005$ and several values of $R$. The black line is the Schwarzschild case.}
\label{fig:lapse}
\end{figure}

Because $M_{\rm DM}>0$, the lapse function stays below the Schwarzschild one (Fig.~\ref{fig:lapse}), and the horizon, given by $r_h=2[M+M_{\rm DM}(r_h)]$, lies outside $r=2M$. For $\rho_0=0.005$, for instance, $r_h$ grows from $2.0178$ at $R=0.5$ to $2.2141$ at $R=2$ (Table~\ref{tab:eik}).

Since $M_{\rm DM}$ grows logarithmically, the total mass is not finite and the spacetime has no ADM mass. This is a consequence of the $r^{-3}$ fall off; for the Hernquist profile, which decreases as $r^{-4}$, the total mass is finite~\cite{Qi2026}. The metric nevertheless approaches the Minkowski form, since $1-f=O(\ln r/r)$. The tortoise coordinate behaves as $r_*=r+\mathcal{O}(\ln^2 r)$, the potentials derived below decay as $\ell(\ell+1)/r^2$ up to logarithmic corrections, and the usual outgoing wave condition can be imposed at infinity. In a realistic situation the $r^{-3}$ fall off is cut off at some outer radius (King's original profile contains a tidal radius~\cite{King1962}), which makes the total mass finite without affecting the region near the BH. We return to this point in Sec.~\ref{sec:ecking}.

Some care is needed when interpreting the numbers. Rotation curves of dark matter dominated galaxies point to cores of roughly constant density with sizes of the order of kiloparsecs~\cite{deBlok2010}, and the dark matter density in the solar neighbourhood is about $10^{-24}\,{\rm g\,cm^{-3}}$~\cite{Read2014}. For a BH with the mass of Sgr~A$^*$ this gives $\rho_0M^2\lesssim10^{-26}$ in geometric units, and the halo mass inside the photon sphere is smaller than $M$ by more than twenty orders of magnitude. The values used below, $\rho_0=0.0025$ to $0.01$ and $R=0.5$ to $2$, describe instead a compact core of a size comparable to the horizon, for which $M_{\rm DM}(r_{ph})$ ranges from about $1\%$ to $21\%$ of $M$. They explore the strong coupling regime, in the same spirit as Ref.~\cite{Hassanabadi2026}, where $M/R=1$ was used. As shown in Sec.~\ref{sec:meff}, the frequencies are controlled mainly by the enclosed mass, so the results can be scaled to other parameter values.

\section{Perturbations and WKB computation}
\label{sec:pert}
\begin{figure}[htbp]
\centering
\includegraphics[width=\textwidth]{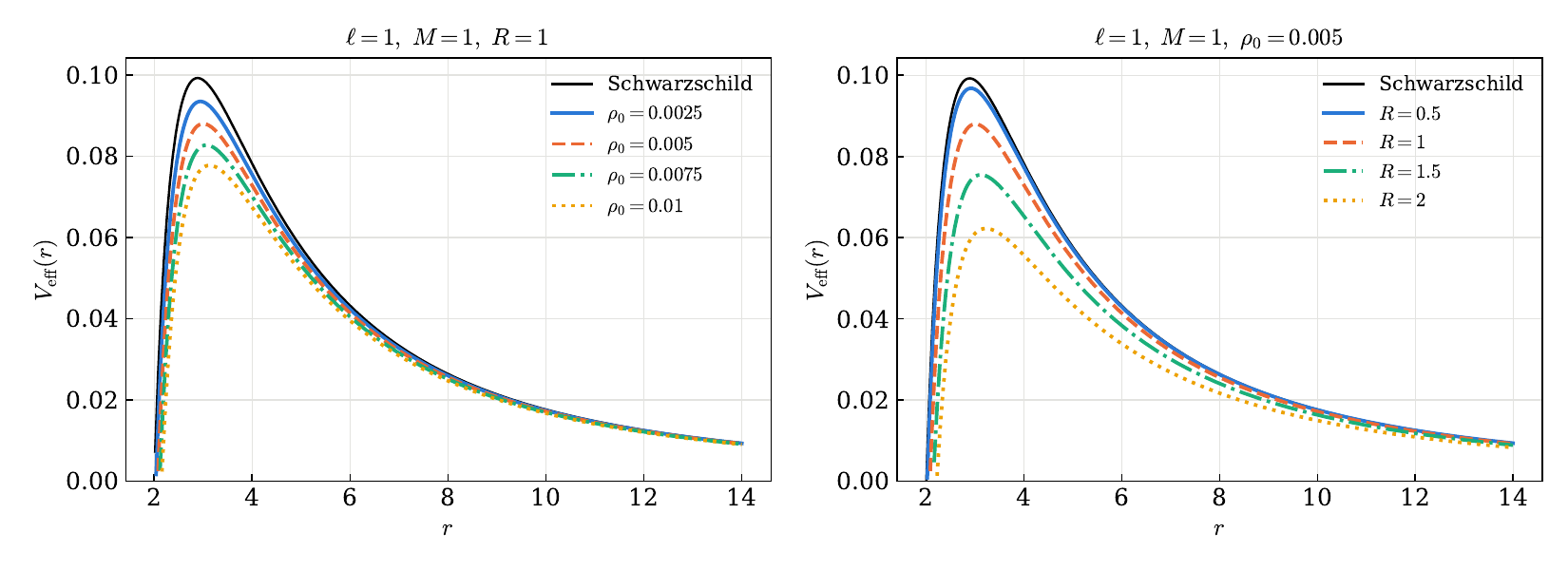}
\caption{Scalar effective potential for $\ell=1$. Left: $R=1$ and several values of $\rho_0$. Right: $\rho_0=0.005$ and several values of $R$. The black line is the Schwarzschild potential.}
\label{fig:Vr}
\end{figure}

We now formulate the scalar and electromagnetic wave equations and outline the higher-order WKB procedure used to estimate their quasinormal frequencies. A massless scalar field obeys $\partial_\mu(\sqrt{-g}\,g^{\mu\nu}\partial_\nu\Phi)=0$. With $\Phi=e^{-i\omega t}\,Y_{\ell m}(\theta,\varphi)\,\psi(r)/r$ and the tortoise coordinate $dr_*=dr/f$, the radial equation takes the form
\begin{equation}
\frac{d^2\psi}{dr_*^2}+\left[\omega^2-V(r)\right]\psi=0 .
\label{eq:schr}
\end{equation}
For metrics with $g_{tt}g_{rr}=-1$ the electromagnetic field leads to the same equation~\cite{Konoplya2011}, and the two potentials can be written together as
\begin{equation}
V(r)=f(r)\left[\frac{\ell(\ell+1)}{r^2}+\left(1-s^2\right)\frac{f'(r)}{r}\right],
\label{eq:V}
\end{equation}
with $s=0$ for the scalar and $s=1$ for the electromagnetic field ($\ell\ge1$). The halo enters through $f$ and, for the scalar field, through $f'=2(M+M_{\rm DM})/r^2-8\pi r\rho$. The potential vanishes at the horizon and asymptotically at infinity, forming a single barrier between these limits (Fig.~\ref{fig:Vr}). This barrier structure is the condition that makes the standard higher-order WKB treatment applicable to the modes considered below. The QNMs are the solutions that behave as $e^{-i\omega r_*}$ for $r_*\to-\infty$ and as $e^{+i\omega r_*}$ for $r_*\to+\infty$.

For a single-peaked potential barrier, the QNM frequencies can be estimated using the WKB condition introduced by Schutz and Will~\cite{Schutz1985}, extended to third order by Iyer and Will~\cite{Iyer1987} and to sixth order by Konoplya~\cite{Konoplya2003},
\begin{equation}
\frac{i\left(\omega^2-V_0\right)}{\sqrt{-2V_0''}}-\sum_{k=2}^{6}\Lambda_k=n+\frac12 ,
\label{eq:wkb}
\end{equation}
where $V_0$ and $V_0''$ are the value of the potential and of its second $r_*$ derivative at the maximum, and $\Lambda_k$ depends on derivatives of $V$ up to order $2k$. We used the implementation of Ref.~\cite{Konoplya2019} and report both the sixth order value and the average of the Pad\'e approximants of the same series~\cite{Matyjasek2017}.

The sixth-order formula requires derivatives of $V$ through twelfth order, and the representation in Eq.~\eqref{eq:f} causes numerical difficulties. Every derivative of $\ln(\sqrt{r^2+R^2}-r)$ produces another factor $1/(\sqrt{r^2+R^2}-r)$. Near the peak this quantity is small (about $0.16$ for $r\simeq3$, $R=1$), and the individual terms of the high derivatives become much larger than their sum. In double precision the third order result was still reasonable, but the fourth order correction was already wrong. The problem disappears if $f$ is written with ${\rm arcsinh}(r/R)$, whose derivatives are algebraic, and if the WKB formulas are applied to a local expansion of the potential. We located the maximum $r_0$ with 50 digit precision, computed the derivatives of $f$ and $V$ at $r_0$ up to sixteenth order, and passed the resulting Taylor polynomials in $x=r-r_0$ to the WKB routines. For $\rho_0=0$ this gives $0.292910-0.097762\,i$ ($\ell=1$) and $0.483642-0.096766\,i$ ($\ell=2$) for the scalar field, the known sixth order values~\cite{Konoplya2003}. An independent code based on the same formulas reproduces all sixth order frequencies to better than $10^{-6}$.

\section{Quasinormal spectrum}
\label{sec:results}
The resulting frequencies are presented below, with particular attention to the fundamental modes and to the dependence on the halo parameters.

\begin{table}[htbp]\centering\small\setlength{\tabcolsep}{4pt}
\caption{Scalar quasinormal frequencies for $\ell=1$ ($M=1$). For every $(\rho_0,R)$ the first line is the sixth order WKB value and the second line the Pad\'e-averaged value of the same series. The fundamental mode ($n=0$) is the main focus of the analysis. The overtones are included for completeness; in particular, the condition $n<\ell$ is not satisfied for $n=1,2$ when $\ell=1$, so these modes should be interpreted with additional caution.}
\label{tab:l1}
\begin{tabular}{llllll}\hline\hline
$\rho_0$ & $R$ & & $\omega_0$ & $\omega_1$ & $\omega_2$\\\hline
0 & -- & WKB$_6$ & $0.292910-0.097762\,i$ & $0.264471-0.306518\,i$ & $0.231014-0.542165\,i$\\
 & & Pad\'e & $0.292931-0.097661\,i$ & $0.264456-0.306507\,i$ & $0.228850-0.542165\,i$\\
\hline
0.005 & 0.5 & WKB$_6$ & $0.289429-0.096240\,i$ & $0.261566-0.301644\,i$ & $0.228673-0.533347\,i$\\
 & & Pad\'e & $0.289448-0.096145\,i$ & $0.261551-0.301633\,i$ & $0.226567-0.533344\,i$\\
0.005 & 1 & WKB$_6$ & $0.276750-0.090129\,i$ & $0.251228-0.281992\,i$ & $0.220384-0.497696\,i$\\
 & & Pad\'e & $0.276765-0.090055\,i$ & $0.251218-0.281983\,i$ & $0.218672-0.497640\,i$\\
0.005 & 1.5 & WKB$_6$ & $0.257590-0.080703\,i$ & $0.235212-0.251873\,i$ & $0.206499-0.443132\,i$\\
 & & Pad\'e & $0.257660-0.080573\,i$ & $0.235114-0.251654\,i$ & $0.206118-0.442971\,i$\\
0.005 & 2 & WKB$_6$ & $0.235034-0.070180\,i$ & $0.215050-0.217682\,i$ & $0.184830-0.378525\,i$\\
 & & Pad\'e & $0.235244-0.069840\,i$ & $0.214794-0.216905\,i$ & $0.184629-0.378360\,i$\\
\hline
0.0025 & 1 & WKB$_6$ & $0.284778-0.093892\,i$ & $0.257817-0.294077\,i$ & $0.225651-0.519603\,i$\\
 & & Pad\'e & $0.284796-0.093805\,i$ & $0.257805-0.294067\,i$ & $0.223734-0.519569\,i$\\
0.0075 & 1 & WKB$_6$ & $0.268831-0.086474\,i$ & $0.244706-0.270271\,i$ & $0.215206-0.476462\,i$\\
 & & Pad\'e & $0.268844-0.086416\,i$ & $0.244697-0.270264\,i$ & $0.213662-0.476395\,i$\\
0.01 & 1 & WKB$_6$ & $0.261027-0.082932\,i$ & $0.238253-0.258923\,i$ & $0.210109-0.455919\,i$\\
 & & Pad\'e & $0.261036-0.082887\,i$ & $0.238244-0.258916\,i$ & $0.208699-0.455848\,i$\\
\hline\hline\end{tabular}\end{table}

\begin{table}[htbp]\centering\small\setlength{\tabcolsep}{4pt}
\caption{Same as Table~\ref{tab:l1} for $\ell=2$.}
\label{tab:l2}
\begin{tabular}{llllll}\hline\hline
$\rho_0$ & $R$ & & $\omega_0$ & $\omega_1$ & $\omega_2$\\\hline
0 & -- & WKB$_6$ & $0.483642-0.096766\,i$ & $0.463847-0.295627\,i$ & $0.430386-0.508700\,i$\\
 & & Pad\'e & $0.483644-0.096759\,i$ & $0.463846-0.295626\,i$ & $0.430330-0.508664\,i$\\
\hline
0.005 & 0.5 & WKB$_6$ & $0.477931-0.095273\,i$ & $0.458554-0.291021\,i$ & $0.425774-0.500643\,i$\\
 & & Pad\'e & $0.477933-0.095267\,i$ & $0.458554-0.291019\,i$ & $0.425719-0.500608\,i$\\
0.005 & 1 & WKB$_6$ & $0.457293-0.089296\,i$ & $0.439653-0.272524\,i$ & $0.409636-0.468148\,i$\\
 & & Pad\'e & $0.457294-0.089291\,i$ & $0.439652-0.272523\,i$ & $0.409586-0.468117\,i$\\
0.005 & 1.5 & WKB$_6$ & $0.426602-0.080030\,i$ & $0.411367-0.243836\,i$ & $0.384986-0.417694\,i$\\
 & & Pad\'e & $0.426603-0.080026\,i$ & $0.411365-0.243829\,i$ & $0.384967-0.417676\,i$\\
0.005 & 2 & WKB$_6$ & $0.391161-0.069616\,i$ & $0.377805-0.211580\,i$ & $0.353650-0.360860\,i$\\
 & & Pad\'e & $0.391168-0.069602\,i$ & $0.377798-0.211541\,i$ & $0.353635-0.360837\,i$\\
\hline
0.0025 & 1 & WKB$_6$ & $0.470392-0.092980\,i$ & $0.451694-0.283912\,i$ & $0.419978-0.488123\,i$\\
 & & Pad\'e & $0.470393-0.092974\,i$ & $0.451693-0.283911\,i$ & $0.419925-0.488090\,i$\\
0.0075 & 1 & WKB$_6$ & $0.444354-0.085716\,i$ & $0.427732-0.261469\,i$ & $0.399365-0.448785\,i$\\
 & & Pad\'e & $0.444355-0.085713\,i$ & $0.427731-0.261468\,i$ & $0.399318-0.448755\,i$\\
0.01 & 1 & WKB$_6$ & $0.431586-0.082242\,i$ & $0.415941-0.250753\,i$ & $0.389171-0.430041\,i$\\
 & & Pad\'e & $0.431586-0.082240\,i$ & $0.415940-0.250752\,i$ & $0.389104-0.430001\,i$\\
\hline\hline\end{tabular}\end{table}

\begin{table}[htbp]\centering\small\setlength{\tabcolsep}{4pt}
\caption{Electromagnetic quasinormal frequencies for $\ell=1$ and $\ell=2$ ($M=1$). For every $(\rho_0,R)$ the first line is the sixth order WKB value and the second line the Pad\'e-averaged value.}
\label{tab:em}
\begin{tabular}{lllcccc}\hline\hline
 & & & \multicolumn{2}{c}{$\ell=1$} & \multicolumn{2}{c}{$\ell=2$}\\
$\rho_0$ & $R$ & & $\omega_0$ & $\omega_1$ & $\omega_0$ & $\omega_1$\\\hline
0 & -- & WKB$_6$ & $0.248191-0.092637\,i$ & $0.214295-0.294118\,i$ & $0.457593-0.095011\,i$ & $0.436534-0.290728\,i$\\
 & & Pad\'e & $0.248251-0.092485\,i$ & $0.214271-0.294105\,i$ & $0.457595-0.095005\,i$ & $0.436533-0.290727\,i$\\
\hline
0.005 & 0.5 & WKB$_6$ & $0.245458-0.091244\,i$ & $0.212268-0.289556\,i$ & $0.452317-0.093563\,i$ & $0.431706-0.286247\,i$\\
 & & Pad\'e & $0.245515-0.091099\,i$ & $0.212244-0.289543\,i$ & $0.452319-0.093557\,i$ & $0.431705-0.286246\,i$\\
0.005 & 1 & WKB$_6$ & $0.235926-0.085700\,i$ & $0.205640-0.271239\,i$ & $0.433507-0.087790\,i$ & $0.414761-0.268327\,i$\\
 & & Pad\'e & $0.235959-0.085609\,i$ & $0.205603-0.271224\,i$ & $0.433507-0.087786\,i$ & $0.414760-0.268326\,i$\\
0.005 & 1.5 & WKB$_6$ & $0.221954-0.077152\,i$ & $0.195756-0.243178\,i$ & $0.405826-0.078858\,i$ & $0.389704-0.240579\,i$\\
 & & Pad\'e & $0.221983-0.077086\,i$ & $0.195733-0.243149\,i$ & $0.405827-0.078856\,i$ & $0.389703-0.240578\,i$\\
0.005 & 2 & WKB$_6$ & $0.205601-0.067568\,i$ & $0.182920-0.211619\,i$ & $0.374002-0.068790\,i$ & $0.360051-0.209310\,i$\\
 & & Pad\'e & $0.205735-0.067357\,i$ & $0.182736-0.211220\,i$ & $0.374007-0.068780\,i$ & $0.360046-0.209289\,i$\\
\hline
0.0025 & 1 & WKB$_6$ & $0.242045-0.089124\,i$ & $0.209985-0.282517\,i$ & $0.445497-0.091353\,i$ & $0.425616-0.279375\,i$\\
 & & Pad\'e & $0.242090-0.089005\,i$ & $0.209957-0.282504\,i$ & $0.445498-0.091349\,i$ & $0.425615-0.279374\,i$\\
0.0075 & 1 & WKB$_6$ & $0.229839-0.082367\,i$ & $0.201264-0.260288\,i$ & $0.421634-0.084323\,i$ & $0.403978-0.257590\,i$\\
 & & Pad\'e & $0.229863-0.082299\,i$ & $0.201213-0.260272\,i$ & $0.421634-0.084320\,i$ & $0.403977-0.257588\,i$\\
0.01 & 1 & WKB$_6$ & $0.223791-0.079126\,i$ & $0.196861-0.249671\,i$ & $0.409888-0.080955\,i$ & $0.393277-0.247171\,i$\\
 & & Pad\'e & $0.223809-0.079080\,i$ & $0.196757-0.249649\,i$ & $0.409889-0.080953\,i$ & $0.393273-0.247168\,i$\\
\hline\hline\end{tabular}\end{table}

The scalar frequencies are listed in Tables~\ref{tab:l1} and~\ref{tab:l2} and the electromagnetic ones in Table~\ref{tab:em}, for $\rho_0=0.005$ with $R=0.5$ to $2$ and for $R=1$ with $\rho_0=0.0025$ to $0.01$. For the fundamental modes the sixth order and Pad\'e values differ by at most $4\times10^{-4}$ for $\ell=1$ and $4\times10^{-5}$ for $\ell=2$. In the discussion we use the Pad\'e values.

\begin{figure}[htbp]
\centering
\includegraphics[width=\textwidth]{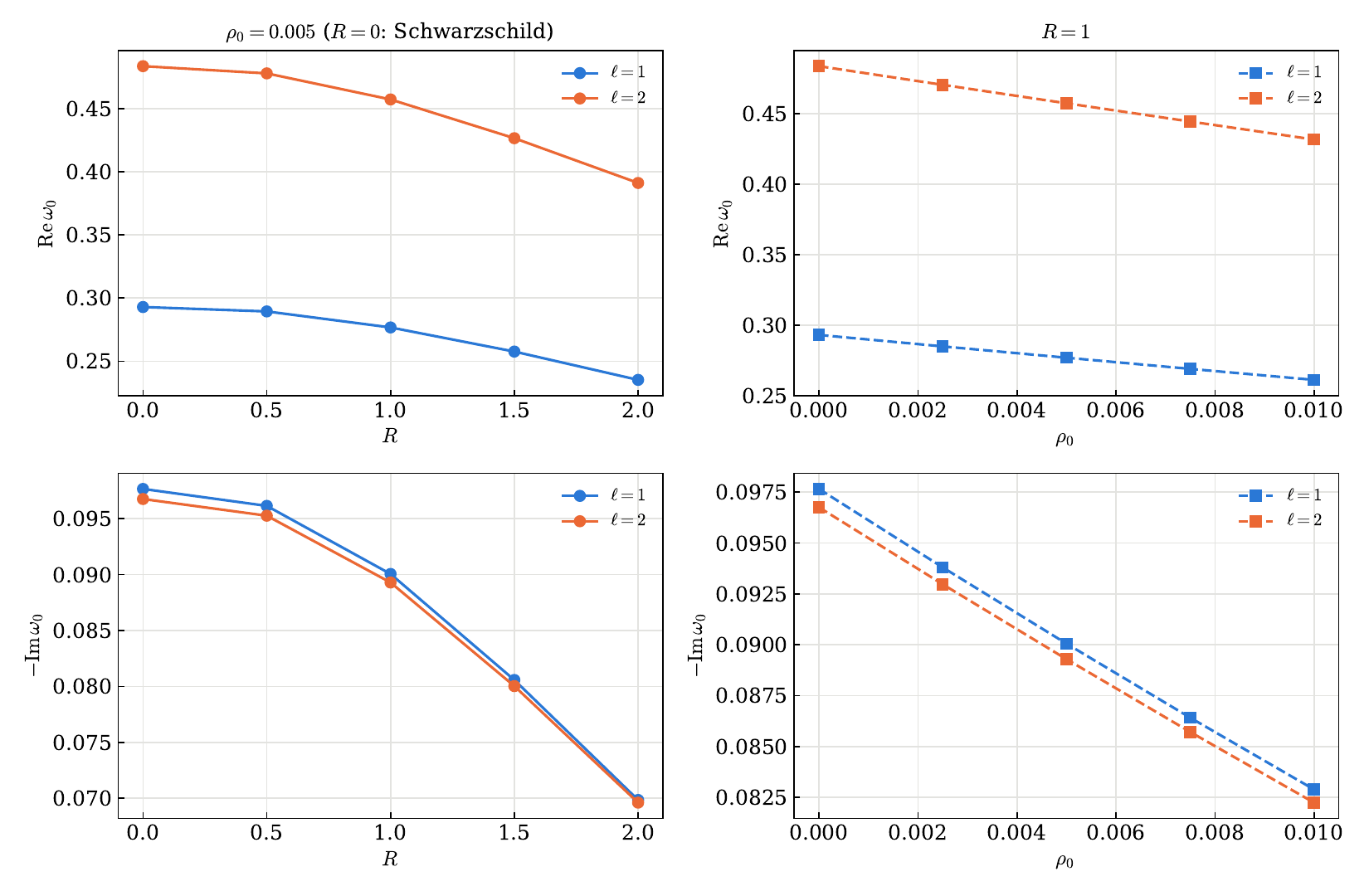}
\caption{Real part (top) and minus the imaginary part (bottom) of the fundamental scalar mode (Pad\'e values). Left: $\rho_0=0.005$ as a function of $R$ (the point at $R=0$ is the Schwarzschild value). Right: $R=1$ as a function of $\rho_0$.}
\label{fig:omega0}
\end{figure}

The halo lowers ${\rm Re}\,\omega$ and $|{\rm Im}\,\omega|$ for all modes and both fields, and the effect grows monotonically with $\rho_0$ and $R$ (Fig.~\ref{fig:omega0}). For $\rho_0=0.005$ and $R=0.5$ the fundamental $\ell=2$ scalar mode changes by $1.2\%$ in the real part and $1.5\%$ in the damping rate, whereas for $R=2$ the changes reach $19\%$ and $28\%$. At fixed $R=1$ the shift grows roughly linearly with $\rho_0$. The relative shifts hardly depend on $\ell$ or on the field: for $\rho_0=0.005$, $R=2$ the real part decreases by $19.7\%$ for $\ell=1$ and $19.1\%$ for $\ell=2$, and the electromagnetic $\ell=2$ mode changes by $18\%$ and $28\%$. This suggests that the halo acts mostly through an overall rescaling of the frequency. All frequencies have ${\rm Im}\,\omega<0$, in qualitative agreement with Ref.~\cite{Hassanabadi2026}.

\subsection{Effective mass and the circular null orbit}
\label{sec:meff}
To interpret the frequency shifts, we compare the halo-supported spectrum with the Schwarzschild spectrum at an effective mass and with the properties of the circular null orbit. Since the halo enters the metric only through $M_{\rm DM}(r)$, one may ask whether the frequencies are those of a heavier Schwarzschild BH. The natural candidate for the effective mass is the mass inside the photon sphere, $M_{\rm eff}=M+M_{\rm DM}(r_{ph})$, because the low multipoles are generated near the maximum of the potential. Table~\ref{tab:meff} shows the test for the scalar $\ell=2$ mode. The rescaled real part ${\rm Re}\,\omega_0M_{\rm eff}$ differs from its Schwarzschild value by less than $0.3\%$ for $R\le1$ and by less than $2\%$ for $R=2$, where ${\rm Re}\,\omega_0$ itself has decreased by $19\%$. For $\ell=1$ the deviations do not exceed $2.6\%$, and the electromagnetic modes stay within $0.9\%$. The oscillation frequency is therefore fixed, to a good approximation, by the dark matter enclosed by the photon sphere.

The damping rate does not follow this rule; after the rescaling it is still $13\%$ below the vacuum value for $R=2$. For large $\ell$ both potentials reduce to $\ell(\ell+1)f/r^2$, whose maximum lies on the circular null orbit $rf'=2f$, and~\cite{Cardoso2009}
\begin{equation}
\omega\simeq\left(\ell+\frac12\right)\Omega_c-i\left(n+\frac12\right)\lambda,\qquad
\Omega_c=\frac{\sqrt{f(r_{ph})}}{r_{ph}},\qquad
\lambda=\sqrt{\frac{f(r_{ph})\left[2f(r_{ph})-r_{ph}^2f''(r_{ph})\right]}{2r_{ph}^2}},
\label{eq:eik}
\end{equation}
where $\Omega_c$ and $\lambda$ are the angular velocity and the Lyapunov exponent of the orbit. For a Schwarzschild BH $\lambda/\Omega_c=1$ for any mass, so this ratio isolates the part of the halo effect that cannot be absorbed into $M$. The last column of Table~\ref{tab:meff} shows that it follows the rescaled imaginary part closely. Because $M_{\rm DM}(r)$ is still growing at $r_{ph}$, the peak of the potential is flatter than for a point mass with the same height, the orbit is less unstable, and the perturbation leaks out more slowly.

\begin{table}[htbp]
\centering\small
\caption{Halo mass inside the photon sphere and rescaled fundamental frequency of the scalar $\ell=2$ mode (Pad\'e values). $\mathcal{R}_{\rm Re}$ and $\mathcal{R}_{\rm damp}$ are the real part and positive damping rate $-\operatorname{Im}\omega_0$, respectively, multiplied by $M_{\rm eff}$ and divided by their Schwarzschild values; a value of one would mean that the shift is entirely due to $M_{\rm eff}=M+M_{\rm DM}(r_{ph})$.}
\label{tab:meff}
\begin{tabular}{cccccccc}
\hline\hline
$\rho_0$ & $R$ & $8\pi\rho_0R^3$ & $M_{\rm DM}(r_{ph})$ & $M_{\rm eff}$ & $\mathcal{R}_{\rm Re}$ & $\mathcal{R}_{\rm damp}$ & $\lambda/\Omega_c$\\
\hline
0.005 & 0.5 & 0.0157 & 0.0119 & 1.0119 & 0.9999 & 0.9963 & 0.9964\\
0.005 & 1   & 0.1257 & 0.0567 & 1.0567 & 0.9991 & 0.9751 & 0.9760\\
0.005 & 1.5 & 0.4241 & 0.1277 & 1.1277 & 0.9947 & 0.9327 & 0.9376\\
0.005 & 2   & 1.0053 & 0.2134 & 1.2134 & 0.9814 & 0.8728 & 0.8891\\
\hline
0.0025 & 1  & 0.0628 & 0.0278 & 1.0278 & 0.9997 & 0.9876 & 0.9880\\
0.0075 & 1  & 0.1885 & 0.0867 & 1.0867 & 0.9984 & 0.9626 & 0.9642\\
0.01   & 1  & 0.2513 & 0.1178 & 1.1178 & 0.9975 & 0.9500 & 0.9526\\
\hline\hline
\end{tabular}
\end{table}

\begin{table}[htbp]\centering\small\setlength{\tabcolsep}{7pt}
\caption{Horizon radius $r_h$, photon sphere radius $r_{ph}$, critical impact parameter $b_c=1/\Omega_c$, and $\Omega_c$ and $\lambda$ of the circular null orbit ($M=1$). The last two columns compare the eikonal estimate $(\ell+\tfrac12)\Omega_c-\tfrac{i}{2}\lambda$ for $\ell=2$ with the Pad\'e-averaged WKB value.}
\label{tab:eik}
\resizebox{\textwidth}{!}{\begin{tabular}{ccccccccc}\hline\hline
$\rho_0$ & $R$ & $r_h$ & $r_{ph}$ & $b_c$ & $\Omega_c$ & $\lambda$ & $\omega^{\rm eik}_{\ell=2}$ & $\omega_0$ (WKB, $\ell=2$)\\\hline
0 & -- & 2.00000 & 3.00000 & 5.19615 & 0.19245 & 0.19245 & $0.481125-0.096225\,i$ & $0.483644-0.096759\,i$\\
0.005 & 0.5 & 2.01779 & 3.02814 & 5.25800 & 0.19019 & 0.18949 & $0.475466-0.094747\,i$ & $0.477933-0.095267\,i$\\
0.005 & 1 & 2.07224 & 3.11584 & 5.49326 & 0.18204 & 0.17768 & $0.455103-0.088840\,i$ & $0.457294-0.089291\,i$\\
0.005 & 1.5 & 2.14210 & 3.22492 & 5.88145 & 0.17003 & 0.15942 & $0.425065-0.079709\,i$ & $0.426603-0.080026\,i$\\
0.005 & 2 & 2.21415 & 3.32395 & 6.39801 & 0.15630 & 0.13896 & $0.390747-0.069481\,i$ & $0.391168-0.069602\,i$\\
0.0025 & 1 & 2.03530 & 3.05651 & 5.34137 & 0.18722 & 0.18497 & $0.468045-0.092483\,i$ & $0.470393-0.092974\,i$\\
0.0075 & 1 & 2.11092 & 3.17816 & 5.65214 & 0.17692 & 0.17060 & $0.442310-0.085298\,i$ & $0.444355-0.085713\,i$\\
0.01 & 1 & 2.15144 & 3.24365 & 5.81831 & 0.17187 & 0.16372 & $0.429678-0.081859\,i$ & $0.431586-0.082240\,i$\\
\hline\hline\end{tabular}}\end{table}

The quantities of the null orbit are listed in Table~\ref{tab:eik}. The critical impact parameter and the photon sphere grow with both halo parameters; for $\rho_0=0.005$, $R=2$ we find $b_c=6.398$, compared with $3\sqrt3\simeq5.196$ in vacuum. For the parameter sets listed in Table~\ref{tab:eik}, the eikonal formula~\eqref{eq:eik} differs from the Pad\'e-averaged WKB value by about $0.1\%$--$0.5\%$ for $\ell=2$. This comparison indicates that the unstable null orbit captures much of the spectral dependence on $(\rho_0,R)$ in these cases.

\section{Time domain analysis}
\label{sec:td}
We independently test the frequency estimates by evolving the wave equation in the time domain and extracting the dominant ringdown oscillations.

As an independent check, we integrated the time-dependent wave equation in the light-cone coordinates $u=t-r_*$ and $v=t+r_*$,
\begin{equation}
4\frac{\partial^2\psi}{\partial u\,\partial v}+V\psi=0 ,
\label{eq:uv}
\end{equation}
with the scheme of Gundlach, Price and Pullin~\cite{Gundlach1994}. The function $r(r_*)$ was obtained by integrating $dr/dr_*=f$ numerically, with $r_*=0$ at $r=3$. The initial data were a Gaussian pulse $\psi(0,v)=\exp[-(v-20)^2/8]$, the domain was $0\le u,v\le500$ with step $h=0.1$, and the signal was recorded at $r_*=10$. The dominant frequency was extracted from the ringdown with the matrix pencil method~\cite{Hua1990}, using $65\le t\le135$ for $\ell=1$ and $60\le t\le150$ for $\ell=2$.

\begin{table}[htbp]\centering\small\setlength{\tabcolsep}{8pt}
\caption{Relative difference $\delta=|\omega^{\rm TD}-\omega^{\rm WKB}|/|\omega^{\rm WKB}|$, in per cent, between the fundamental mode extracted from the time domain signals and the sixth order WKB value ($M=1$, $\rho_0=0.005$ except for the Schwarzschild row).}
\label{tab:td}
\begin{tabular}{ccccc}\hline\hline
$R$ & scalar, $\ell=1$ & scalar, $\ell=2$ & EM, $\ell=1$ & EM, $\ell=2$\\\hline
Schw. & 0.043 & 0.015 & 0.087 & 0.004\\
0.5 & 0.042 & 0.011 & 0.087 & 0.004\\
1 & 0.042 & 0.005 & 0.074 & 0.003\\
1.5 & 0.083 & 0.001 & 0.058 & 0.002\\
2 & 0.198 & 0.006 & 0.160 & 0.005\\
\hline\hline\end{tabular}\end{table}

\begin{figure}[htbp]
\centering
\includegraphics[width=\textwidth]{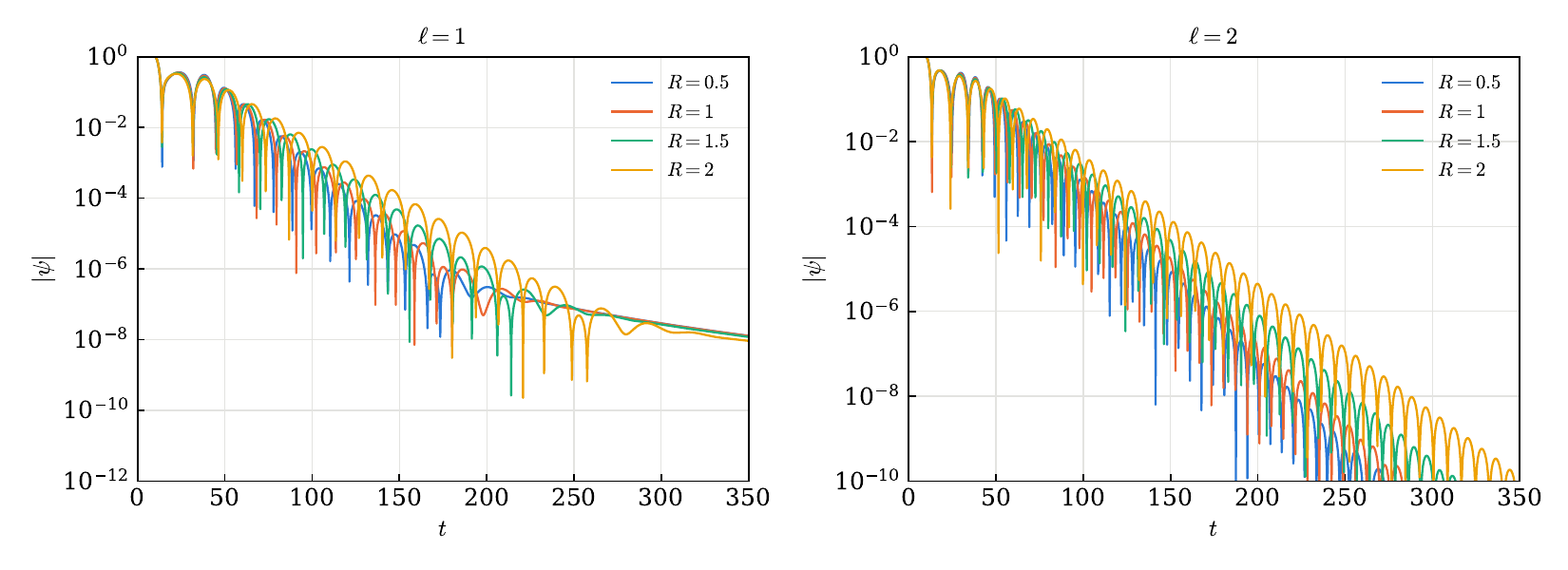}
\caption{Time domain profiles of the scalar field on a logarithmic scale for $M=1$, $\rho_0=0.005$ and $R=0.5,1,1.5,2$. Left: $\ell=1$. Right: $\ell=2$.}
\label{fig:td_log}
\end{figure}

\begin{figure}[htbp]
\centering
\includegraphics[width=\textwidth]{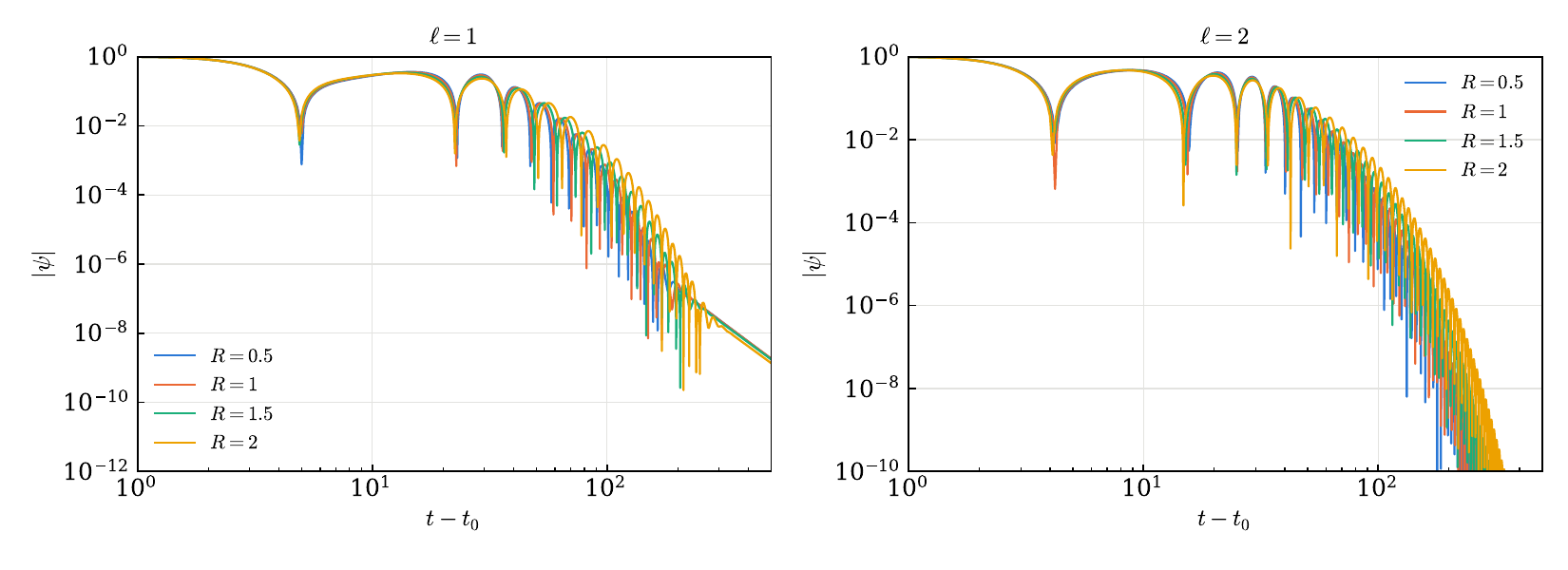}
\caption{Double logarithmic plot of $|\psi|$ versus $t-t_0$ ($t_0$ is the first recorded time) for the signals of Fig.~\ref{fig:td_log}.}
\label{fig:td_loglog}
\end{figure}

Table~\ref{tab:td} shows that the time-domain and WKB frequencies agree to within $0.02\%$ for $\ell=2$ and $0.2\%$ for $\ell=1$ for the configurations examined. These differences quantify the agreement between the two methods; a separate convergence study with respect to the integration step, extraction window, and matrix-pencil fitting parameters is needed to quantify the associated numerical uncertainties. Figure~\ref{fig:td_log} shows the scalar signals for $\rho_0=0.005$; larger cores ring more slowly and for longer. At late times the ringdown is replaced by a power law tail (Fig.~\ref{fig:td_loglog}). For $\ell=1$ a fit of $\ln|\psi|$ against $\ln t$ on $250<t<480$ gives exponents between $-5.26$ and $-4.94$, and the Schwarzschild signal analysed in the same way gives $-5.26$.  The fitted exponents are consistent, over the selected time window, with the expected $t^{-(2\ell+3)}$ behavior~\cite{Price1972}. No discernible change in the fitted exponent is seen over this window; this finite-time fit does not by itself establish the strict asymptotic tail.

\section{A King halo outside the horizon}
\label{sec:ec}
We next replace the phenomenological source extending through the horizon with an Einstein-cluster description in which the halo can be confined to the exterior region. In the metric of Sec.~\ref{sec:geometry}, part of the dark matter lies inside the event horizon, where it cannot represent a population of static halo particles. For $f=1-2m(r)/r$ the field equations require an anisotropic fluid with $\rho=m'/(4\pi r^2)$, radial pressure $P_r=-\rho$ and tangential pressure $P_t=-\rho-r\rho'/2$. For the King profile $-\rho\leq P_t\leq\rho/2$, so outside the horizon the null, weak and dominant energy conditions hold, while the strong energy condition is violated for $r<\sqrt2\,R$. Inside the horizon $T^t{}_t=T^r{}_r=-\rho$, so the stress tensor in the $(t,r)$ block is proportional to the identity and the energy density measured by any observer is at least $\rho>0$; the weak energy condition therefore holds there as well. The difficulty of the model is thus not a negative energy, but the interpretation of its source. The radial pressure is a tension equal in magnitude to the energy density, and the profile does not follow from a microscopic model of the dark matter. This motivates a construction in which the halo is made of collisionless matter and lies outside the BH.

\subsection{Einstein cluster}
The Einstein-cluster construction provides the metric and matter conditions needed to describe collisionless particles on circular orbits around the BH. We consider
\begin{equation}
ds^2=-a(r)\,dt^2+\frac{dr^2}{b(r)}+r^2\left(d\theta^2+\sin^2\theta\,d\varphi^2\right),\qquad b(r)=1-\frac{2m(r)}{r},
\label{eq:ecmetric}
\end{equation}
sourced by particles moving on circular geodesics with all orientations, the Einstein cluster~\cite{Einstein1939}. The radial pressure vanishes, the tangential pressure is $P_t=m\rho/[2(r-2m)]$, and the field equations give $a'/a=2m/[r(r-2m)]$. With $m(r)=M$ near the horizon and $a\to1$ at infinity,
\begin{equation}
a(r)=\left(1-\frac{2M}{r}\right)e^{\Upsilon(r)},\qquad
\Upsilon(r)=-\int_r^\infty\frac{2\left[m(s)-M\right]}{\left[s-2m(s)\right]\left(s-2M\right)}\,ds .
\label{eq:ups}
\end{equation}
The horizon stays at $r=2M$. Where $m=M$ the metric is the Schwarzschild one with the time rescaled by $e^{\Upsilon/2}$, which is the gravitational redshift produced by the halo between a given radius and infinity. A consistent cluster requires $r>2m(r)$ for $r>2M$, a nondecreasing $m(r)$ and timelike orbits, $v^2=m/(r-2m)<1$, i.e.
\begin{equation}
r>3m(r)\qquad\text{wherever}\ \rho(r)>0 .
\label{eq:vlimit}
\end{equation}
In the vacuum Schwarzschild region adjacent to the horizon, stable circular orbits begin at the innermost stable circular orbit, $r=6M$. This condition is distinct from the requirement $r>3m(r)$ for a timelike circular orbit and from the dominant-energy-condition bound $2r\geq5m(r)$. The same thresholds, $r_{\rm in}/r_h\geq5/4$, $r_{\rm in}/r_h>3/2$ and $r_{\rm in}/r_h\geq3$, were found for Dehnen halos in Ref.~\cite{Hassanabadi2026b}; the first had been obtained earlier by Shen, Wang and Yin~\cite{Shen2025} for a Hernquist halo with an inner edge.

\subsection{King profile outside the horizon}
\label{sec:ecking}
We implement the exterior King halo using two mass profiles, distinguishing a smooth density that vanishes at the horizon from a halo with a prescribed inner edge. Following Ref.~\cite{Cardoso2022} we first take
\begin{equation}
m_A(r)=M+M_{\rm DM}(r)\left(1-\frac{2M}{r}\right)^2,
\label{eq:famA}
\end{equation}
for which the density vanishes linearly at the horizon and approaches the King density far away (family A). In family B the density is the King density for $r\geq r_{\rm in}$ and zero inside,
\begin{equation}
m_B(r)=M+\Theta(r-r_{\rm in})\left[M_{\rm DM}(r)-M_{\rm DM}(r_{\rm in})\right],
\label{eq:famB}
\end{equation}
so an empty region separates the horizon from the halo. In Ref.~\cite{Hassanabadi2026b} the mass inside $r_{\rm in}$ is absorbed into the BH; here it is removed, so the horizon stays at $2M$.

The density of family B jumps at $r_{\rm in}$. For the Einstein cluster this needs no surface layer. The junction conditions~\cite{Israel1966} require the induced metric and the extrinsic curvature to be continuous, and the relevant components of the latter are proportional to $\sqrt b/r$ and $\sqrt b\,a'/a$, which depend only on $m$. Since the radial pressure vanishes on both sides, only $m(r)$ has to be continuous. For a metric with $g_{tt}g_{rr}=-1$, by contrast, $P_r=-\rho$ would jump with the density and a thin shell would be needed.

For both families $m(r)$ still grows logarithmically, but the integrand in Eq.~\eqref{eq:ups} decreases as $(\ln s)/s^2$ and $\Upsilon$ is finite. A truncation at a large radius $r_t$, with the Schwarzschild metric of mass $m(r_t)$ outside, makes the total mass finite and the spacetime asymptotically flat, and again requires no surface layer. For $\rho_0=0.005$, $R=2$ the redshift factor at the photon sphere of family A is $0.80616$ without truncation and $0.80699$ for $r_t=500$; for $r_t\ge500$ the frequencies thus change by about $0.1\%$ or less.

\begin{figure}[htbp]
\centering
\includegraphics[width=\textwidth]{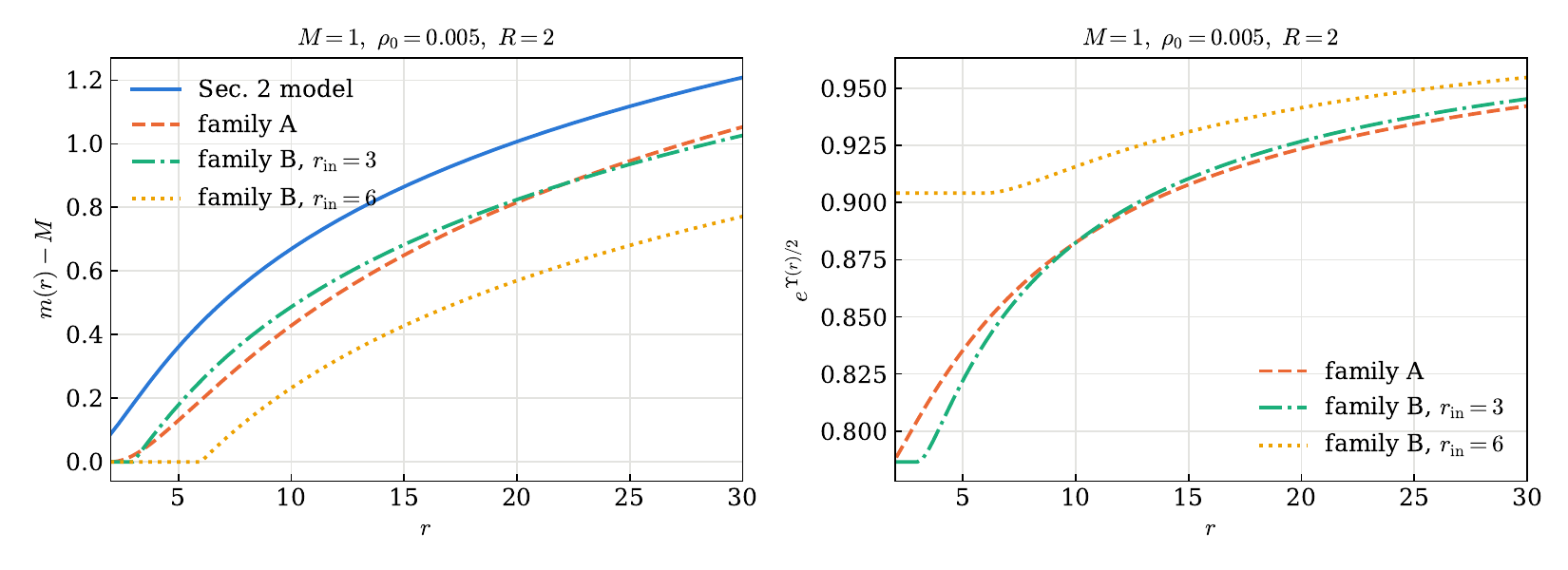}
\caption{Left: halo mass $m(r)-M$ for $M=1$, $\rho_0=0.005$, $R=2$ in the model of Sec.~\ref{sec:geometry}, in family A and in family B with $r_{\rm in}=3$ and $6$. Right: redshift factor $e^{\Upsilon(r)/2}$ for the three outside halos.}
\label{fig:ecmass}
\end{figure}

The mass functions and the redshift factor are shown in Fig.~\ref{fig:ecmass}. For all our parameter sets no second horizon appears. In family A the condition~\eqref{eq:vlimit} fails between $r=2M$ and the radius where $r=3m_A$; the matter there is small, $0.0013M$ to $0.023M$, but the cluster interpretation does not hold in this region, and for $2M<r<5m_A/2$ the dominant energy condition is also violated. In family B with $r_{\rm in}\geq3M$ all conditions are satisfied. The null, weak and strong energy conditions hold for both families. The physical consistency of the cluster therefore favours an empty region between the horizon and the halo.

\subsection{Perturbations and the test with a Hernquist halo}
We derive the corresponding perturbation potentials and check the numerical implementation against a Hernquist-halo case with published reference frequencies. The perturbations satisfy Eq.~\eqref{eq:schr} with $dr_*=dr/\sqrt{ab}$ and
\begin{equation}
V_{\rm EM}=a\,\frac{\ell(\ell+1)}{r^2},\qquad
V_{\rm s}=a\,\frac{\ell(\ell+1)}{r^2}+\frac{1}{2r}\,\frac{d(ab)}{dr}.
\label{eq:ecV}
\end{equation}
The maximum of $V_{\rm EM}$ lies exactly on the circular null orbit $r=3m(r)$. For the WKB method only $a(r_0)$ has to be computed numerically; its higher Taylor coefficients follow from $a'/a=2m/[r(r-2m)]$ by the recursion $(k+1)\alpha_{k+1}=\sum_{j=0}^k\gamma_j\alpha_{k-j}$, where $\alpha_k$ and $\gamma_k$ are the coefficients of $a$ and of $2m/[r(r-2m)]$ in powers of $r-r_0$. In the time domain we integrate $r$ and $\Upsilon$ along $r_*$, using $\ln(r-2M)$ as the variable near the horizon.

\begin{table}[htbp]\centering\small\setlength{\tabcolsep}{4pt}
\caption{Fundamental electromagnetic modes of the BH with a Hernquist halo, Eq.~\eqref{eq:hern}, in units $M_{\rm BH}=1/2$: our Pad\'e-averaged WKB and time domain values compared with those of Ref.~\cite{Konoplya2021}.}
\label{tab:hern}
\begin{tabular}{cccccc}\hline\hline
$\ell$ & $a_0/M_{\rm BH}$ & $M_{\rm h}/M_{\rm BH}$ & Pad\'e & time domain & Ref.~\cite{Konoplya2021}\\\hline
1 & 1000 & 100 & $0.447848-0.166786\,i$ & $0.44786-0.16676\,i$ & $0.447853-0.166808\,i$\\
1 & 100 & 30 & $0.359372-0.132574\,i$ & $0.35944-0.13249\,i$ & $0.359376-0.132591\,i$\\
1 & 400 & 200 & $0.272649-0.101143\,i$ & $0.27278-0.10096\,i$ & $0.272652-0.101157\,i$\\
2 & 1000 & 100 & $0.825492-0.171333\,i$ & $0.82549-0.17132\,i$ & $0.825493-0.171334\,i$\\
2 & 100 & 30 & $0.662078-0.136236\,i$ & $0.66207-0.13619\,i$ & $0.662079-0.136237\,i$\\
2 & 400 & 200 & $0.502460-0.103916\,i$ & $0.50245-0.10391\,i$ & $0.502461-0.103916\,i$\\
\hline\hline\end{tabular}\end{table}

We tested the method on the Hernquist model of Ref.~\cite{Cardoso2022},
\begin{equation}
m(r)=M_{\rm BH}+\frac{M_{\rm h}\,r^2}{(a_0+r)^2}\left(1-\frac{2M_{\rm BH}}{r}\right)^2,
\label{eq:hern}
\end{equation}
for which $\Upsilon$ is known in closed form~\cite{Konoplya2021}; our numerical integration reproduces it to about $10^{-13}$. Table~\ref{tab:hern} compares our frequencies with the higher order WKB values of Konoplya~\cite{Konoplya2021}. The Pad\'e-averaged values agree to within $2.2\times10^{-5}$ for $\ell=1$ and $2\times10^{-6}$ for $\ell=2$, and the time domain values agree to the accuracy of the fit.

\subsection{Quasinormal spectrum}
We now compare the fundamental modes of the exterior-halo models and separate the effects associated with gravitational redshift from those due to the shape of the effective potential.

\begin{table}[htbp]\centering\small\setlength{\tabcolsep}{4pt}
\caption{Fundamental modes of family A, Eq.~\eqref{eq:famA}, Pad\'e-averaged WKB values ($M=1$). The first line is the Schwarzschild case.}
\label{tab:ecA}
\begin{tabular}{cccccc}\hline\hline
 & & \multicolumn{2}{c}{scalar} & \multicolumn{2}{c}{electromagnetic}\\
$\rho_0$ & $R$ & $\ell=1$ & $\ell=2$ & $\ell=1$ & $\ell=2$\\\hline
0 & -- & $0.29293-0.09766\,i$ & $0.48364-0.09676\,i$ & $0.24825-0.09248\,i$ & $0.45759-0.09501\,i$\\\hline
0.005 & 0.5 & $0.29104-0.09672\,i$ & $0.48050-0.09583\,i$ & $0.24678-0.09164\,i$ & $0.45470-0.09411\,i$\\
0.005 & 1 & $0.28187-0.09248\,i$ & $0.46534-0.09165\,i$ & $0.23953-0.08774\,i$ & $0.44066-0.09004\,i$\\
0.005 & 1.5 & $0.26337-0.08461\,i$ & $0.43494-0.08388\,i$ & $0.22467-0.08035\,i$ & $0.41235-0.08243\,i$\\
0.005 & 2 & $0.23555-0.07377\,i$ & $0.38936-0.07312\,i$ & $0.20193-0.06993\,i$ & $0.36969-0.07181\,i$\\
0.0025 & 1 & $0.28738-0.09506\,i$ & $0.47447-0.09419\,i$ & $0.24388-0.09010\,i$ & $0.44911-0.09251\,i$\\
0.0075 & 1 & $0.27638-0.08993\,i$ & $0.45626-0.08913\,i$ & $0.23519-0.08540\,i$ & $0.43225-0.08760\,i$\\
0.01 & 1 & $0.27093-0.08741\,i$ & $0.44723-0.08664\,i$ & $0.23087-0.08308\,i$ & $0.42388-0.08518\,i$\\
\hline\hline\end{tabular}\end{table}

The fundamental modes of family A are listed in Table~\ref{tab:ecA}; the sixth order values differ from these by less than $4\times10^{-5}$ for $\ell=2$ and $5\times10^{-4}$ for $\ell=1$. For $\rho_0=0.005$, $R=2$ the scalar $\ell=2$ mode decreases by $19.5\%$ in the real part and $24.4\%$ in the damping rate, close to the $19\%$ and $28\%$ of Sec.~\ref{sec:results}, although the halo mass inside the photon sphere is now about ten times smaller ($0.023M$ against $0.213M$).

\begin{table}[htbp]\centering\small\setlength{\tabcolsep}{5pt}
\caption{Family A: redshift factor at the photon sphere, ratios $\mathcal{R}_{\rm Re}$ and $\mathcal{R}_{\rm damp}$ of the scalar $\ell=2$ oscillation frequency and positive damping rate, respectively, to their Schwarzschild values}, and $\lambda/\Omega_c$. The last three columns give the redshift factor $e^{\Upsilon(r_{\rm in})/2}$ of family B, by which the sixth order WKB frequencies of family B are the Schwarzschild ones multiplied.
\label{tab:ecratio}
\begin{tabular}{cccccccccc}\hline\hline
 & & \multicolumn{5}{c}{family A} & \multicolumn{3}{c}{family B}\\
$\rho_0$ & $R$ & $e^{\Upsilon(r_{ph})/2}$ & $\mathcal{R}_{\rm Re}$ & $\mathcal{R}_{\rm damp}$ & $\mathcal{R}_{\rm damp}/\mathcal{R}_{\rm Re}$ & $\lambda/\Omega_c$ & $r_{\rm in}=4$ & $r_{\rm in}=6$ & $r_{\rm in}=10$\\\hline
0.005 & 0.5 & 0.9935 & 0.9935 & 0.9904 & 0.9969 & 0.9969 & 0.9973 & 0.9984 & 0.9991\\
0.005 & 1 & 0.9622 & 0.9622 & 0.9472 & 0.9845 & 0.9846 & 0.9790 & 0.9875 & 0.9930\\
0.005 & 1.5 & 0.8996 & 0.8993 & 0.8670 & 0.9641 & 0.9642 & 0.9322 & 0.9585 & 0.9766\\
0.005 & 2 & 0.8062 & 0.8050 & 0.7560 & 0.9390 & 0.9387 & 0.8487 & 0.9042 & 0.9451\\
0.0025 & 1 & 0.9810 & 0.9810 & 0.9735 & 0.9923 & 0.9924 & 0.9895 & 0.9937 & 0.9965\\
0.0075 & 1 & 0.9434 & 0.9434 & 0.9212 & 0.9765 & 0.9767 & 0.9685 & 0.9812 & 0.9895\\
0.01 & 1 & 0.9248 & 0.9247 & 0.8955 & 0.9684 & 0.9687 & 0.9581 & 0.9749 & 0.9861\\
\hline\hline\end{tabular}\end{table}

Table~\ref{tab:ecratio} explains the size of the shift. The ratio of the real parts to the Schwarzschild ones coincides with the redshift factor $e^{\Upsilon(r_{ph})/2}$ to within $1.2\times10^{-3}$, and the extra reduction of the damping rate, $\mathcal{R}_{\rm damp}/\mathcal{R}_{\rm Re}$, agrees to within $3\times10^{-4}$ with $\lambda/\Omega_c$, where now $\Omega_c=\sqrt{a(r_{ph})}/r_{ph}$ and $\lambda^2=-\tfrac12r_{ph}^2b(r_{ph})\,(a/r^2)''_{r_{ph}}$. The real part is thus the Schwarzschild value seen through the redshift of the halo, and the damping rate carries in addition the flattening of the barrier. In this model the redshift produced by the whole halo plays the role that the enclosed mass played in Sec.~\ref{sec:meff}. For family B with $r_{\rm in}>3M$ the barrier lies in the empty region, and the WKB frequencies are exactly the Schwarzschild ones multiplied by $e^{\Upsilon(r_{\rm in})/2}$, in agreement with the eikonal result of Ref.~\cite{Hassanabadi2026b}. For $\rho_0=0.005$, $R=2$ they are reduced by $15\%$, $10\%$ and $5\%$ for $r_{\rm in}=4$, $6$ and $10$.

\subsection{Scattering by the inner edge}
\label{sec:ectdsec}
Because the WKB approximation is local to the potential maximum, we use time-domain evolution and reflection coefficients to assess additional scattering from the halo's inner edge. The WKB result uses only the potential near its maximum and cannot see waves reflected by the halo outside the barrier. We therefore studied family B, and family A for comparison, in the time domain and through the reflection coefficient. For family B the time domain frequencies are compared with the Schwarzschild time domain values multiplied by $e^{\Upsilon(r_{\rm in})/2}$, so that small errors of the fit cancel; in the Schwarzschild limit our code reproduces the scalar $\ell=1$ and $\ell=2$ modes to about $10^{-5}$. The scalar potential contains $b'$ and is discontinuous at $r_{\rm in}$, while the electromagnetic one has only a jump in its second derivative.

\begin{table}[htbp]\centering\small\setlength{\tabcolsep}{5pt}
\caption{Relative differences, in per cent, of the real part and of the damping rate between the time domain frequencies and the reference values, for $\rho_0=0.005$, $R=2$. The reference is the Pad\'e-averaged WKB value for family A and $e^{\Upsilon(r_{\rm in})/2}\omega^{\rm S}$ for family B.}
\label{tab:ectd}
\begin{tabular}{lcccccccc}\hline\hline
 & \multicolumn{2}{c}{scalar, $\ell=1$} & \multicolumn{2}{c}{scalar, $\ell=2$} & \multicolumn{2}{c}{EM, $\ell=1$} & \multicolumn{2}{c}{EM, $\ell=2$}\\
case & Re & Im & Re & Im & Re & Im & Re & Im\\\hline
A & $-0.01$ & $-0.10$ & $0.00$ & $0.00$ & $0.00$ & $-0.05$ & $0.00$ & $-0.01$\\
B, $r_{\rm in}=3$ & $-0.08$ & $-8.95$ & $-0.03$ & $-8.70$ & $+1.52$ & $-8.16$ & $+0.47$ & $-8.46$\\
B, $r_{\rm in}=4$ & $-1.46$ & $-1.60$ & $-0.68$ & $-0.02$ & $-0.65$ & $-3.25$ & $-0.50$ & $-1.20$\\
B, $r_{\rm in}=6$ & $-0.03$ & $+2.32$ & $+0.22$ & $+0.67$ & $-0.36$ & $+0.58$ & $+0.05$ & $+0.58$\\
B, $r_{\rm in}=10$ & $+0.08$ & $-1.29$ & $-0.09$ & $+0.55$ & $+0.12$ & $-0.05$ & $-0.03$ & $+0.03$\\
\hline\hline\end{tabular}\end{table}

The results are given in Table~\ref{tab:ectd}. For family A the two methods agree to within $0.1\%$. For family B with $r_{\rm in}=3$ the edge sits on the top of the barrier, and the damping rates are $8\%$ to $9\%$ smaller than the redshifted Schwarzschild values. For $r_{\rm in}=4$ the differences fall to at most $1.5\%$ in the real part and $3.3\%$ in the damping rate, and for $r_{\rm in}\ge6$ they stay below $0.7\%$, except for the damping rate of the scalar dipole mode (up to $2.3\%$), which has the broadest barrier and a discontinuous potential at the edge.

\begin{figure}[htbp]
\centering
\includegraphics[width=\textwidth]{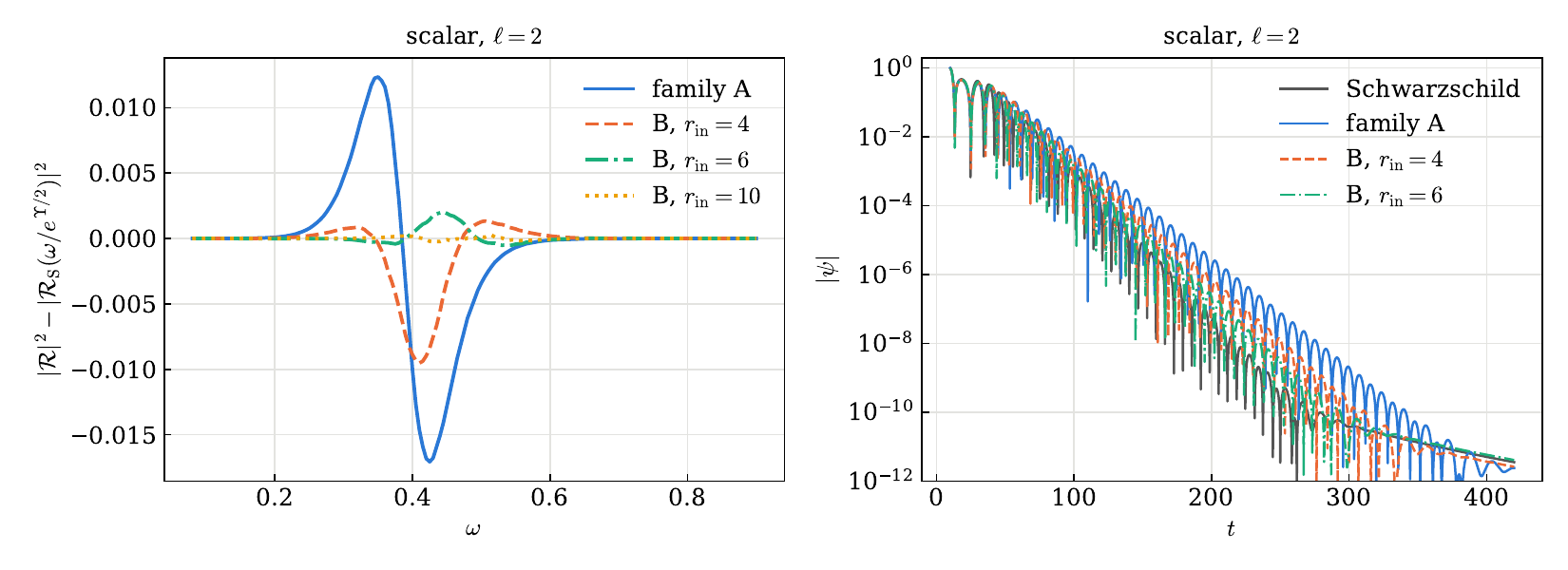}
\caption{Scalar $\ell=2$ perturbations for $\rho_0=0.005$, $R=2$. Left: reflection coefficient minus $|\mathcal{R}_{\rm S}(\omega/e^{\Upsilon/2})|^2$, with the redshift factor at the photon sphere (family A) or at the inner edge (family B). Right: time domain signals at $r_*=10$.}
\label{fig:ecscatter}
\end{figure}

The reflection coefficient $|\mathcal{R}(\omega)|^2$ was computed at real frequency with a purely ingoing wave at the horizon; its numerical error, estimated from $|\mathcal{R}|^2+|\mathcal{T}|^2=1$, is about $3\times10^{-4}$. Figure~\ref{fig:ecscatter} shows its difference from the Schwarzschild value at the redshifted frequency, which would vanish if the halo acted only through the redshift. The difference reaches $0.017$ for family A and decreases from $0.010$ to $0.002$ and below $10^{-3}$ for family B with $r_{\rm in}=4$, $6$ and $10$. The curves show no sharp resonances, and the time domain signals show no delayed pulses or echoes within our time window.

\subsection{Shape of the inner edge}
\label{sec:edge}
Finally, we vary the regularity of the density profile at the inner edge to determine how smoothing the transition changes wave scattering and the resulting frequencies. To see how much of this scattering comes from the jump of the density, we repeated the calculation for the profiles
\begin{equation}
\rho_n(r)=\rho(r)\left(1-\frac{r_{\rm in}}{r}\right)^{n}\Theta(r-r_{\rm in}),\qquad n=0,1,2,
\label{eq:rhon}
\end{equation}
taken from the family of Eq.~(4.2) of Ref.~\cite{Hassanabadi2026b}. The case $n=0$ is family B. For $n=1$ the density is continuous and its derivative jumps; this choice was used in Ref.~\cite{Hassanabadi2026b} because in the Hernquist case it contains the solution of Cardoso \emph{et al.}~\cite{Cardoso2022} for $r_{\rm in}=r_h$ and the profile of Shen, Wang and Yin~\cite{Shen2025}. The case $n=2$, with a continuous derivative as well, serves only as a test. The mass inside $r_{\rm in}$ is again removed, and $m(r)$ follows in closed form from the primitives of $r^2\rho$, $r\rho$ and $\rho$.

\begin{table}[htbp]\centering\small\setlength{\tabcolsep}{6pt}
\caption{Halos with the edge profile~\eqref{eq:rhon} for $\rho_0=0.005$, $R=2$. Columns four and five give the largest relative difference, in per cent, between the time domain frequencies and $e^{\Upsilon(r_{\rm in})/2}\omega^{\rm S}$ over the scalar and electromagnetic modes with $\ell=1,2$; the last column gives the largest deviation of the scalar $\ell=2$ reflection coefficient from $|\mathcal{R}_{\rm S}(\omega/e^{\Upsilon/2})|^2$.}
\label{tab:edge}
\begin{tabular}{cccccc}\hline\hline
$r_{\rm in}$ & $n$ & $e^{\Upsilon(r_{\rm in})/2}$ & $\max|\delta_{\rm Re}|$ & $\max|\delta_{\rm Im}|$ & $\max\big|\delta|\mathcal{R}|^2\big|$\\\hline
3 & 0 & $0.7866$ & $1.52$ & $8.95$ & $0.0211$\\
 & 1 & $0.9010$ & $0.41$ & $1.47$ & $0.0034$\\
 & 2 & $0.9364$ & $0.16$ & $0.32$ & $0.0010$\\\hline
4 & 0 & $0.8487$ & $1.46$ & $3.25$ & $0.0095$\\
 & 1 & $0.9284$ & $0.21$ & $0.46$ & $0.0007$\\
 & 2 & $0.9536$ & $0.06$ & $0.24$ & $0.0002$\\\hline
6 & 0 & $0.9042$ & $0.36$ & $2.32$ & $0.0020$\\
 & 1 & $0.9540$ & $0.10$ & $0.18$ & $0.0004$\\
 & 2 & $0.9700$ & $0.03$ & $0.04$ & $0.0003$\\\hline
10 & 0 & $0.9451$ & $0.12$ & $1.29$ & $0.0003$\\
 & 1 & $0.9733$ & $0.05$ & $0.09$ & $0.0003$\\
 & 2 & $0.9825$ & $0.01$ & $0.02$ & $0.0003$\\
\hline\hline\end{tabular}\end{table}

Table~\ref{tab:edge} shows that the departures from the redshift rule fall rapidly with $n$. At $r_{\rm in}=3$ the largest change of the damping rate is $9\%$, $1.5\%$ and $0.3\%$ for $n=0$, $1$ and $2$, and at $r_{\rm in}=4$ it is $3.3\%$, $0.5\%$ and $0.24\%$. The smoothing also removes matter near the edge, but this alone does not explain the numbers: a sharp edge at $r_{\rm in}=10$, with redshift factor $0.945$, still changes the damping rate by $1.3\%$, whereas the halo with $n=1$ and $r_{\rm in}=4$, which has more matter near the barrier and a smaller factor ($0.928$), gives at most $0.46\%$. The scattering found for family B is therefore caused mainly by the jump of the density, and for a continuous density with the edge outside the barrier the frequencies follow the redshift rule to within about half a per cent.

\section{Conclusions}
\label{sec:conc}

We have computed scalar and electromagnetic QNMs for a Schwarzschild BH embedded in a King dark matter halo, using the sixth-order WKB formula, its Pad\'e-improved version, and time-domain evolution. Expressing the lapse function through the enclosed halo mass improves the numerical stability of the high-order WKB calculation and makes the physical role of the halo more transparent. The halo lowers both the oscillation frequency and the damping rate. The real part of the fundamental mode is, to within a few per cent, that of a Schwarzschild BH with the mass enclosed by the photon sphere, and the additional reduction of the damping rate is measured by the ratio $\lambda/\Omega_c$ of the circular null orbit. For the configurations examined, the WKB and time-domain fundamental frequencies agree to within $0.2\%$. No unstable modes are found within the perturbation sectors and parameter configurations examined. The scalar $\ell=1$ tail is consistent with the $t^{-(2\ell+3)}$ law over the time window analyzed.

The first model should therefore be regarded as a phenomenological reference: its metric is an exact solution with a positive energy density, but the source has a radial tension and extends inside the horizon. Since part of the halo lies inside the horizon in this metric, we also placed the King halo outside the BH as an Einstein cluster. A consistent cluster requires the halo to begin outside the circular null orbit, so a model with an empty region around the horizon is preferred. The frequencies then decrease by a similar amount, but for a different reason: the real part equals the Schwarzschild value multiplied by the redshift factor of the halo at the photon sphere, to within $10^{-3}$, and the damping rate carries in addition the factor $\lambda/\Omega_c$. The halo outside the barrier also scatters the waves. A jump in the density at the inner edge changes the damping rate by up to $9\%$ when the edge lies on the barrier. For a continuous density profile, such as the edge profile of Ref.~\cite{Hassanabadi2026b}, deviations from the redshift relation remain below $1.5\%$ at $r_{\rm in}=3M$ and below $0.5\%$ for $r_{\rm in}\geq4M$. No echoes were found for the dilute halos considered here.

For realistic galactic halo densities, the effects considered here are expected to be extremely small. They become appreciable in the parameter regime studied only when the dark matter forms a compact, dense core whose characteristic size is comparable to the horizon scale. Natural extensions include gravitational perturbations with a consistent treatment of halo dynamics, a broader range of halo compactness, and other cored or cuspy density profiles. These conclusions are limited to linear test-field perturbations, the static spherical backgrounds specified here, and the numerical parameter ranges reported in the tables. A self-consistent treatment of gravitational ringdown and halo dynamics is needed before extrapolating these results to environmental corrections in realistic gravitational-wave signals.

\section*{Data availability}
The Mathematica notebook used to obtain the results of this paper is available from the corresponding author on reasonable request.

\end{document}